\font\bbb=msbm10 \font\bbs=msbm7                                   %%%
%\def\bbb{\bf} \def\bbs{\bf}                                       %%%
%%%                                                                %%%
%%%                                                                %%%
%%%%%%%%%%%%%%%%%%%%%%%%%%%%%%%%%%%%%%%%%%%%%%%%%%%%%%%%%%%%%%%%%%%%%%
%%%%%%%%%%%%%%%%%%%%%%%%%%%%%%%%%%%%%%%%%%%%%%%%%%%%%%%%%%%%%%%%%%%%%%

\overfullrule=0pt

\def\C{\hbox{\bbb C}}  
  
\def\R{\hbox{\bbb R}}  
\def\Z{\hbox{\bbb Z}} \def\sZ{\hbox{\bbs Z}}

\def\ASENSfour{{\sl Ann.\ Sci.\ \'Ecole Norm.\ Sup.\ (4)}}

\def\IEEETIT{{\sl IEEE Trans.\ Inform.\ Theory}}

\def\MRL{{\sl Math.\ Res.\ Lett.}}

\def\PRA{{\sl Phys.\ Rev.\ A}}

\def\PRL{{\sl Phys.\ Rev.\ Lett.}}
\def\PRSLA{{\sl Proc.\ Roy.\ Soc.\ Lond.\ A}}

\def\UMN{{\sl Uspekhi Mat.\ Nauk}}

\def\kitaev{\hbox{A. Yu.\ Kitaev}}
\def\gottesman{\hbox{D. Gottesman}}
\def\rains{\hbox{E. M. Rains}}
\def\steane{\hbox{A. M. Steane}}

\def\hfb{\hfil\break}

\catcode`@=11
\newskip\ttglue

   \font\ninerm=cmr9    \font\eightrm=cmr8   \font\sixrm=cmr6
  \font\ninebf=cmbx9   \font\eightbf=cmbx8  \font\sixbf=cmbx6
  \font\nineit=cmti9   \font\eightit=cmti8  
  \font\ninesl=cmsl9   \font\eightsl=cmsl8  
  \font\ninemi=cmmi9   \font\eightmi=cmmi8  \font\sixmi=cmmi6

\font\bigten=cmr10 scaled\magstep2 

\def\ninepoint{\def\rm{\fam0\ninerm}%
  \textfont0=\ninerm \scriptfont0=\sixrm
  \textfont1=\ninemi \scriptfont1=\sixmi
  \textfont\itfam=\nineit  \def\it{\fam\itfam\nineit}%
  \textfont\slfam=\ninesl  \def\sl{\fam\slfam\ninesl}%
  \textfont\bffam=\ninebf  \scriptfont\bffam=\sixbf
    \def\bf{\fam\bffam\ninebf}%
  \tt \ttglue=.5em plus.25em minus.15em
  \normalbaselineskip=11pt
  \setbox\strutbox=\hbox{\vrule height8pt depth3pt width0pt}%
  \normalbaselines\rm}

\def\eightpoint{\def\rm{\fam0\eightrm}%
  \textfont0=\eightrm \scriptfont0=\sixrm
  \textfont1=\eightmi \scriptfont1=\sixmi
  \textfont\itfam=\eightit  \def\it{\fam\itfam\eightit}%
  \textfont\slfam=\eightsl  \def\sl{\fam\slfam\eightsl}%
  \textfont\bffam=\eightbf  \scriptfont\bffam=\sixbf
    \def\bf{\fam\bffam\eightbf}%
  \tt \ttglue=.5em plus.25em minus.15em
  \normalbaselineskip=9pt
  \setbox\strutbox=\hbox{\vrule height7pt depth2pt width0pt}%
  \normalbaselines\rm}

\def\sfootnote#1{\edef\@sf{\spacefactor\the\spacefactor}#1\@sf
      \insert\footins\bgroup\eightpoint
      \interlinepenalty100 \let\par=\endgraf
        \leftskip=0pt \rightskip=0pt
        \splittopskip=10pt plus 1pt minus 1pt \floatingpenalty=20000
        \parskip=0pt\smallskip\item{#1}\bgroup\strut\aftergroup\@foot\let\next}
\skip\footins=12pt plus 2pt minus 2pt
\dimen\footins=30pc

\def\ie{{\it i.e.}}
\def\eg{{\it e.g.}}

\def\RP2{\R P^2}

\def\Kitaevtoric{1}
\def\qHamming{2}
\def\Kitaevfault{3}
\def\Preskill{4}
\def\Freedman{5}
\def\Steanesimple{6}
\def\QECCequiv{7}
\def\Rainspoly{8}
\def\Shornine{9}
\def\Gottesmanthesis{10}
\def\five{11}
\def\CSSseven{12}
\def\Kitaevprivate{13}
\def\Berger{14}
\def\inprep{15}

\magnification=1200
\input epsf.tex

\dimen0=\hsize \divide\dimen0 by 13 \dimendef\chasm=0
\dimen1=\hsize \advance\dimen1 by -\chasm \dimendef\usewidth=1
\dimen2=\usewidth \divide\dimen2 by 2 \dimendef\halfwidth=2
\dimen3=\usewidth \divide\dimen3 by 3 \dimendef\thirdwidth=3
\dimen4=\hsize \advance\dimen4 by -\halfwidth \dimendef\secondstart=4
\dimen5=\halfwidth \advance\dimen5 by -10pt \dimendef\indenthalfwidth=5
\dimen6=\thirdwidth \multiply\dimen6 by 2 \dimendef\twothirdswidth=6
\dimen7=\twothirdswidth \divide\dimen7 by 4 \dimendef\qttw=7
\dimen8=\qttw \divide\dimen8 by 4 \dimendef\qqttw=8
\dimen9=\qqttw \divide\dimen9 by 4 \dimendef\qqqttw=9
\dimen10=\chasm \multiply\dimen10 by 10 \dimendef\fivein=10

\parskip=0pt\parindent=0pt

\line{\hfil October 1998}
\line{\hfil quant-ph/9810055}
\vfill
\centerline{\bf\bigten PROJECTIVE PLANE AND PLANAR QUANTUM CODES}
\bigskip\bigskip
\centerline{\bf Michael H. Freedman$^*$ and David A. Meyer$^{\dagger}$}
\bigskip 
\centerline{\sl $^*$Microsoft Research, 
                One Microsoft Way}
\centerline{\sl Redmond, WA 98052}
\centerline{michaelf@microsoft.com}
\medskip
\centerline{\sl $^{\dagger}$Project in Geometry and Physics,
                Department of Mathematics}
\centerline{\sl University of California/San Diego,
                La Jolla, CA 92093-0112}
\centerline{dmeyer@chonji.ucsd.edu}
\smallskip
\centerline{\sl and Institute for Physical Sciences,
                Los Alamos, NM}
\vfill
\centerline{ABSTRACT}
\bigskip
%--------|---------|---------|---------|---------|---------|---------|
\noindent Cellulations of the projective plane $\RP2$ define single
qubit topological quantum error correcting codes since there is a 
unique essential cycle in $H_1(\RP2;\Z_2)$.  We construct three of the
smallest such codes, show they are inequivalent, and identify one of 
them as Shor's original 9 qubit repetition code.  We observe that 
Shor's code can be constructed in a planar domain and generalize to
planar constructions of higher genus codes for multiple qubits.

\bigskip\bigskip
%--------|---------|---------|---------|---------|---------|---------|
\noindent PACS numbers:  03.67.Lx,  % Quantum computation
                         89.70.+c,  % Information science
                         89.80.+h.  % Computer science and technology

\noindent AMS subject classification:
                         94Bxx,     % Theory of error correcting codes
                         81P99,     % Quantum theory; foundations
                         57M20.     % Two-dimensional complexes

\noindent KEY WORDS:     topological code, quantum error correction,
                         planar code.
 
\vfill
\eject

\headline{\ninepoint\it Projective plane and planar quantum codes\hfil   
                                                    Freedman \& Meyer}
\parskip=10pt
\parindent=20pt

%--------|---------|---------|---------|---------|---------|---------|
Kitaev has constructed a class of quantum error correcting codes using 
qubits arranged on the edges of square lattices embedded in the two
dimensional torus [\Kitaevtoric].  While these {\sl toric\/} codes are 
not particularly efficient---they do not come close to saturating the 
quantum Hamming bound [\qHamming]---they are nevertheless interesting
for several reasons:  Toric codes have local stabilizers, which means
that the code subspace can be identified as the (degenerate) ground
state subspace of a local Hamiltonian; thus there would be some level 
of automatic error correction in such a quantum system.  Furthermore,
fault-tolerant quantum computation can be performed using elementary 
excitations of the Hamiltonian [\Kitaevfault]; universal quantum 
computation is possible if the qubits (lying in $\C^2 = \C^{\sZ_2}$) 
are replaced in the model with states in $\C^{60} = \C^{A_5}$ 
[\Preskill].

%--------|---------|---------|---------|---------|---------|---------|
Kitaev also remarks that cellulations with $|E|$ edges of genus $g$ 
compact orientable surfaces generally encode $2g$ qubits using $|E|$ 
qubits [\Kitaevfault].  The toric codes, for example, encode 2 qubits.  
In this note we observe that, as is the case for percolation, there is 
something to be learned from studying the problem on the projective 
plane $\RP2$ [\Freedman].  Since there is a unique essential cycle in 
$H_1(\RP2;\Z_2)$, cellulations of $\RP2$ encode a single qubit.  Here 
we consider the smallest such quantum error correcting codes and 
compare them with single qubit codes obtained otherwise.

%--------|---------|---------|---------|---------|---------|---------|
We begin by reviewing the construction of (two dimensional) 
{\sl topological\/} quantum error correcting codes.  A cellulation 
${\cal C}$ of a surface defines sets $F$, $E$ and $V$ of {\sl faces}, 
{\sl edges\/} and {\sl vertices}, respectively.  For each face 
$f \in F$, let $E_f \subset E$ be the set of edges in the boundary of 
$f$; define (our construction is dual to Kitaev's [\Kitaevtoric], but
equivalent)
$$
A_f := \bigotimes_{e\in E} \sigma_x^{\delta(e \in E_f)}.      \eqno(1)
$$
Similarly, for each vertex $v \in V$, let $E_v \subset E$ be the set
of edges in whose boundary $v$ lies; define
$$
B_v := \bigotimes_{e\in E} \sigma_z^{\delta(e \in E_v)}.      \eqno(2)
$$
Here $\sigma_x$ and $\sigma_z$ are the usual Pauli matrices
$$
\sigma_x = \pmatrix{ 0 & 1 \cr
                     1 & 0 \cr
                   }
\quad\hbox{and}\quad
\sigma_z = \pmatrix{ 1 &  0 \cr
                     0 & -1 \cr
                   },
$$
the exponents in equations (1) and (2) are 1(0) according to the 
truth(falsity) of the argument of $\delta(\cdot)$, and the 
{\sl stabilizer\/} operators $A_f$ and $B_v$ act on the Hilbert space 
${\cal H} = (\C^2)^{\otimes|E|}$ with qubit tensor factors labelled by 
the edges of the cellulation.  These stabilizer operators form an 
overcomplete set of generators for the stabilizer group; there are two 
relations:
$$
\prod_f A_f = {\rm id} = \prod_v B_v.                         \eqno(3)
$$

%--------|---------|---------|---------|---------|---------|---------|
As usual, let 0 and 1 denote the eigenvectors of $\sigma_z$ with 
eigenvalues $1$ and $-1$, respectively.  The $2^{|E|}$ configurations 
of 0s and 1s on the edges of ${\cal C}$ form a basis for ${\cal H}$.  
There is a natural bijection between this basis and the set of 
$\Z_2$-linear combinations of elements in $E$, the \hbox{1-chains} of 
${\cal C}$ with coefficients in $\Z_2$, $C_1({\cal C};\Z_2)$; thus we 
identify ${\cal H} = \C^{C_1({\cal C};\sZ_2)}$.  The subspace of 
${\cal H}$ which is the intersection of the eigenvalue 1 eigenspaces 
of the $B_v$ is spanned (over $\C$) by configurations with an even 
number of edges labelled by 1s incident at each vertex; these are 
chains in $C_1({\cal C};\Z_2)$ without boundary, \ie, the cycles 
$Z_1({\cal C};\Z_2)$.  So the subspace of ${\cal H}$ fixed by all the 
$B_v$ is $\C^{Z_1({\cal C};\sZ_2)}$, the functions from $\Z_2$ 
\hbox{1-cycles} to $\C$.  Since $\sigma_x$ acting at an edge exchanges 
0 and 1, each $A_f$ corresponds to the order 2 automorphism of 
$Z_1({\cal C};\Z_2)$ which changes each cycle by the $\Z_2$-addition 
of the cycle bounding $f$.  Thus the subspace of ${\cal H}$ fixed by 
{\sl both\/} all the $B_v$ and all the $A_f$ is the set of functions 
on $Z_1({\cal C};\Z_2)$ which are invariant under the $\Z_2$-addition 
of cycles bounding 2-chains in ${\cal C}$, the boundaries 
$B_1({\cal C};\Z_2)$.  This {\sl code subspace\/} is therefore 
$\C^{H_1({\cal C};\sZ_2)}$, where 
$H_1({\cal C};\Z_2) = Z_1({\cal C};\Z_2)/B_1({\cal C};\Z_2)$ is the 
first homology group of ${\cal C}$ over $\Z_2$.  For a genus $g$ 
compact orientable surface $\Sigma$, 
$H_1(\Sigma;\Z_2) = \bigoplus_{i=1}^{2g} \Z_2$, so a corresponding 
code subspace would have dimension $2^{2g}$ and thus encode $2g$ 
qubits.  But the real projective plane $\RP2$ has a unique essential 
cycle, so $H_1(\RP2;\Z_2) = \Z_2$ and any corresponding code subspace 
has dimension 2 and encodes a single qubit.

%--------|---------|---------|---------|---------|---------|---------|
Consider a bit flip error in the qubit on some edge $e$, \ie, 
multiplication by $\sigma_x$ on the corresponding tensor factor of
${\cal H}$.  This error will be detected by the eigenvalues of the
stabilizer operators $B_v$ for the two vertices in the boundary of 
$e$---unless there are bit flip errors on an even number of the edges
incident at one or the other vertex.  More generally, bit flip errors
in the qubits on any collection of edges corresponding to a chain 
$c_1 \in C_1({\cal C};\Z_2)$ will be detected by the $B_v$ for the 
vertices in the boundary of $c_1$.  Error correction by choosing any 
chain $c_2 \in C_1({\cal C};\Z_2)$ with the observed boundary vertices 
and acting by $\sigma_x$ on the corresponding qubits succeeds unless 
$c_1 + c_2$ contains an essential cycle.  The length of the shortest 
essential cycle in ${\cal C}$ is thus the (bit flip error) distance of 
the code [\Steanesimple].

%--------|---------|---------|---------|---------|---------|---------|
Similarly, phase errors are detected by the eigenvalues of the 
stabilizer operators $A_f$.  The observed faces correspond to vertices 
in the dual cellulation ${\cal C}^*$ bounding a chain 
$c_1^* \in C_1({\cal C}^*;\Z_2)$ of edges at which phase errors 
have occurred (edges in ${\cal C}^*$ are dual to edges in ${\cal C}$).  
Error correction by choosing any $c_2^* \in C_1({\cal C}^*;\Z_2)$ with 
the observed boundary and acting by $\sigma_z$ on the corresponding 
qubits succeeds unless $c_1^* + c_2^*$ contains an essential cycle in 
$C_1({\cal C}^*;\Z_2)$.  The length of the shortest essential cycle in 
the dual cellulation ${\cal C}^*$ is thus the (phase error) distance 
of the code.

%--------|---------|---------|---------|---------|---------|---------|
The smallest square lattice toric code correcting an arbitrary single 
error, \ie, with distance 3, uses 18 qubits to encode two qubits. 
Figure~1 shows a similarly regular triangulation of $\RP2$ with 15 
edges, defining a 15 qubit code for one qubit.  In this diagram 
antipodal points on the circle are identified and it is easy to see 
that while the minimal length of any essential cycle is 3, it is 5 in
the dual cellulation.  By considering more general cellulations we can
construct smaller codes correcting 1 error.  Figure~2 shows a 
cellulation of $\RP2$ with 9 edges, defining a distance 3 code for one 
qubit using 9 qubits; the minimal length of any essential cycle is 3 
in both this cellulation and its dual.  The dual cellulation also 
defines a code, with the stabilizers $A_f$ and $B_v$ replaced by 
corresponding $B_{f^*}$ and $A_{v^*}$.  The resulting code, however, 
is {\sl equivalent\/} [\QECCequiv] to the original under 
multiplication of each tensor factor by the Hadamard transform
$$
H = {1\over\sqrt{2}}\pmatrix{ 1 &  1 \cr
                              1 & -1 \cr
                            }
$$
since $\sigma_z = H\sigma_x H^{-1}$.

%--------|---------|---------|---------|---------|---------|---------|
\topinsert
%\null\vskip-4\baselineskip
$$
\epsfxsize=\thirdwidth\epsfbox{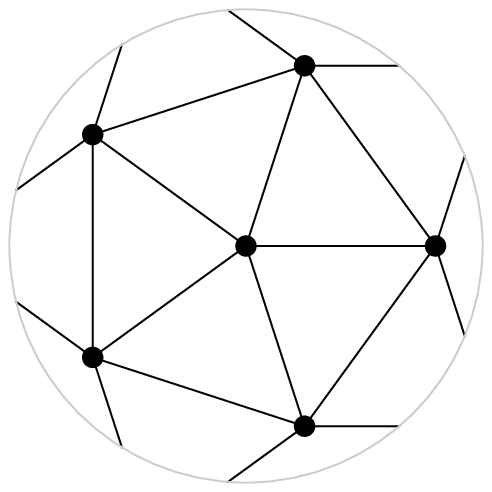}\hskip3\chasm%
\epsfxsize=\thirdwidth\epsfbox{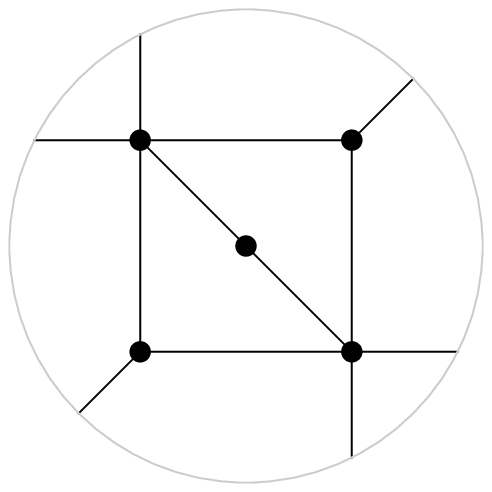}
$$
%\vskip-1.5\baselineskip
\hbox to\hsize{%
\vbox{\hsize=\halfwidth\eightpoint{%
\noindent{\bf Figure~1}.  A triangulation of the projective plane with
15 edges, all minimal essential (dual) cycles of length 3 (5).  
Antipodal points on the circle are identified.
}}
\hfill%
\vbox{\hsize=\halfwidth\eightpoint{%
\noindent{\bf Figure~2}.  A less regular, but more efficient, 
cellulation of the projective plane using 9 edges.  Both the minimal 
essential cycles and the minimal essential dual cycles have length 3.
}}}
\endinsert

%--------|---------|---------|---------|---------|---------|---------|
But there are other cellulations of $\RP2$ which define distinct 
codes.  Figure~3 shows a cellulation obtained from the one in Figure~2
by identifying two vertices.  The resulting 9 qubit code still has 
distance 3 and is not equivalent to the code derived from the 
cellulation in Figure~2.  To demonstrate that these codes are 
inequivalent, we consider the projections $P_1$ and $P_2$ onto the
respective code subspaces---and their polynomial invariants under the
adjoint actions of $U(2)^{\otimes9}$ and the permutation group $S_9$.
There are, of course, many such invariants [\Rainspoly] but for our 
purpose it suffices to consider the coefficients of the characteristic 
polynomials of the reduced density matrices obtained by tracing over 
pairs of tensor factors.  For $P_1$ exactly two of these reduced
density matrices have rank 2 (the two obtained by tracing over either
the qubits on the two edges incident at the valence 2 vertex, or the 
qubits on the edges bounding the 2-gon in the cellulation).  For 
$P_2$, however, there are three rank 2 reduced density 
matrices---corresponding to the presence of the three 2-gons in 
Figure~3.  Thus the two codes are inequivalent.

%--------|---------|---------|---------|---------|---------|---------|
Finally, let us consider the cellulation shown in Figure~4, obtained 
from the one in Figure~3 by identifying two vertices---which 
identifies the endpoints of an edge---and then sliding the endpoints 
of this edge to the two other vertices.  This cellulation again 
defines a 9 qubit code with distance 3.  The resulting code is 
inequivalent to either of our first two by the same argument:  the six 
reduced density matrices corresponding to tracing the projection $P_3$ 
over the qubits on the edges of each 2-gon have rank 2.  While the 
first two codes were new, this code is very familiar---it is exactly 
Shor's original 9 qubit repetition code [\Shornine] as can be seen by 
comparing their stabilizer operators:  The 9 qubits are partitioned 
into 3 triples according to the endpoints of the edges on which they 
lie; there are six stabilizers acting by $\sigma_x$ on pairs of edges 
in the same triple and three stabilizers acting by $\sigma_z$ on all 
the qubits in pairs of triples; one of the latter is redundant, as is 
the stabilizer corresponding to the hexagonal face of the cellulation.  
This is the exactly the (dual of the) stabilizer formulation of Shor's 
code [\Gottesmanthesis, p.\ 17].

%--------|---------|---------|---------|---------|---------|---------|
\topinsert
%\null\vskip-4\baselineskip
$$
\epsfxsize=\thirdwidth\epsfbox{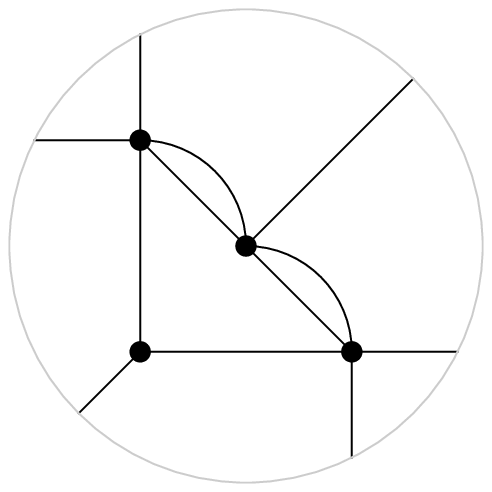}\hskip3\chasm%
\epsfxsize=\thirdwidth\epsfbox{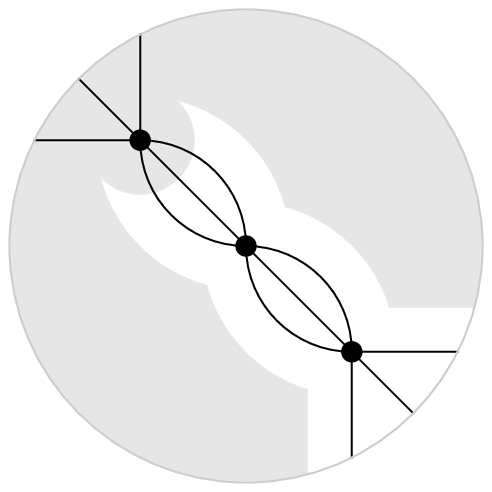}
$$
%\vskip-1.5\baselineskip
\hbox to\hsize{%
\vbox{\hsize=\halfwidth\eightpoint{%
\noindent{\bf Figure~3}.  Another cellulation of the projective plane 
obtained by identifying two vertices of the cellulation shown in
Figure~2.  Both the minimal essential cycles and the minimal essential 
dual cycles still have length 3.
}}
\hfill%
\vbox{\hsize=\halfwidth\eightpoint{%
\noindent{\bf Figure~4}.  A third cellulation of the projective plane 
with minimal essential cycle (also dual cycle) of length 3.  The
interior of one face and a neighborhood of one vertex are shaded; the
remaining cellulation is planar.
}}}
\endinsert

%--------|---------|---------|---------|---------|---------|---------|
By considering cellulations of the projective plane we have 
demonstrated the existence of single qubit topological quantum codes.
While two of the ones we find are new, the third is Shor's original 9
qubit code [\Shornine]; this connects Kitaev's novel perspective
[\Kitaevtoric,\Kitaevfault] with the bulk of the work on quantum error
correcting codes (see, for example, [\Gottesmanthesis] and the 
references therein).  One might ask whether the 5 qubit [\five] and 7 
qubit [\CSSseven] single qubit codes are also equivalent to some 
projective plane quantum code.  They are not---there are no 
cellulations of $\RP2$ with 5 or 7 edges and lengths of all essential 
cycles and dual cycles at least 3.  This exemplifies the inefficiency 
of two dimensional qubit topological quantum error correcting codes, 
even for cellulations with few edges.  We reiterate that their 
attraction lies in the locality of the stabilizer operators which one 
might hope to implement with designer (but local) Hamiltonians. 

%--------|---------|---------|---------|---------|---------|---------|
As Kitaev remarks [\Kitaevfault], for the purposes of physical 
implementation one would like to make two dimensional topological 
quantum error correcting codes {\sl planar\/} in the sense that the 
qubits lie in a plane and that each of the necessary stabilizers acts 
locally, on the qubits at the frontier of one of a set of disjoint 
regions (\eg, a neighborhood of a vertex or the interior of a face).  
Notice that the redundancy of the stabilizer operators implied by the 
relations in equation (3) allows us to disregard one of the faces and
one of the vertices and thus make Shor's code planar:  removing the
interior of the hexagon face and a neighborhood of, say, the upper
left vertex in Figure~4 (both shaded), leaves a planar diagram%
\sfootnote{$^*$}{Perhaps the simplest way to conceptualize the 
                 resulting planarity is to think of the projective 
                 plane as formed by the three faces of a cube incident 
                 at a single vertex with antipodal identification of 
                 the hexagonal boundary.  Discarding two of the three 
                 faces---these represent the domain of the redundant 
                 $A_f$ and $B_v$---the result is a single square with 
                 no boundary identifications, clearly a planar 
                 object.}
with a
generating set of stabilizers acting locally (imagine the qubits to
be located at the midpoints of the edges).

%--------|---------|---------|---------|---------|---------|---------|
\moveright\secondstart\vtop to 0pt{\hsize=\halfwidth
%\vskip -0.5\baselineskip
$$
\epsfxsize=\thirdwidth\epsfbox{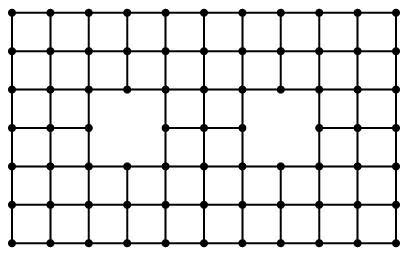}
$$
\vskip 0\baselineskip
\eightpoint{%
\noindent{\bf Figure~5}.  A cellulation of the 2-punctured disk which
defines a planar topological quantum code for two qubits correcting 1 
phase error and 3 bitflip errors.
}}
\vskip -\baselineskip
\parshape=11
0pt \halfwidth
0pt \halfwidth
0pt \halfwidth
0pt \halfwidth
0pt \halfwidth
0pt \halfwidth
0pt \halfwidth
0pt \halfwidth
0pt \halfwidth
0pt \halfwidth
0pt \hsize
%--------|---------|---------|---------|---------|---------|---------|
A related observation leads to planar constructions of topological
quantum error correcting codes for multiple qubits deriving from 
higher genus surfaces [\Kitaevfault]:  the faces of a cellulation need 
not be disks.  For these more general cellulations ${\cal C}$, the 
code subspace corresponds to 
$$
H_1({\cal C};\Z_2)/\bigoplus_f H_1(f,\partial f;\Z_2)
$$
To apply this
observation to construct a planar code protecting $g$ qubits, 
cellulate an orientable surface of genus $g$ using one large face 
with the topology of a $g$-punctured disk and all other faces disks.
Again by the relation (3) we may discard the $A_f$ corresponding to 
the large face; the remaining faces cellulate a $g$-punctured disk
which is, of course, planar.  Particularly simple versions of such
planar codes---with all stabilizers involving no more than 4 
qubits---can be constructed using subsets of the square lattice.
Figure~5 shows such a planar construction for a two qubit topological 
quantum code correcting 1 phase error and 3 bitflip errors.

%--------|---------|---------|---------|---------|---------|---------|
Kitaev and Bravyi have discovered a closely related planar 
construction by a different route [\Kitaevprivate].  Their planar 
lattices have ``$x$-boundary'' and ``$z$-boundary''.  Connecting the 
free edges of the $x$-boundary to an additional vertex (for which the 
associated $B_v$ can be discarded) and taking the $z$-boundary as the 
boundary of an additional face (for which the associated $A_f$ can 
also be discarded) defines a cellulation of a closed surface.  We 
greatly appreciate Alexei Kitaev's willingness to describe their 
preliminary results and his assistance in recognizing the isomorphism 
between their construction and ours.

%--------|---------|---------|---------|---------|---------|---------|
We conclude by remarking that higher dimensional manifolds offer the
possibility of constructing local codes which are more efficient, in
the sense of protecting against more (worst case) errors relative to
their size, than any local surface code.  Their intrinsic geometry 
[\Berger] restricts $n$ qubit surface codes for a constant number of
qubits to correcting $O(n^{1/2})$ (worst case) errors.  But, for 
example, five dimensional $n$ qubit topological quantum codes for a 
constant number of qubits can correct $O(n^{32/61})$ (worst case) 
errors [\inprep].

\vfill\eject

\medskip
\noindent{\bf Acknowledgements}
\nobreak

\nobreak
%--------|---------|---------|---------|---------|---------|---------|
\noindent We thank Alexei Kitaev, Greg Kuperberg, John Preskill, 
Melanie Quong and Nolan Wallach for useful conversations.  This work 
has been partially supported by U. S. Army Research Office grant 
DAAG55-98-1-0376.

\bigskip
\global\setbox1=\hbox{[00]\enspace}
\parindent=\wd1

\noindent{\bf References}
\vskip10pt

\parskip=0pt
%--------|---------|---------|---------|---------|---------|---------|
\item{[\Kitaevtoric]}
\kitaev,
``Quantum computations:  algorithms and error correction'',
\UMN\ {\bf 52} (1997) 53--112.

\item{[\qHamming]}
A. Ekert and C. Macchiavello,
``Error correction in quantum communication'',
\PRL\ {\bf 77} (1996) 2585--2588.

\item{[\Kitaevfault]}
\kitaev,
``Fault-tolerant quantum computation by anyons'',
quant-ph/9707021.

\item{[\Preskill]}
J. Preskill,
``Fault-tolerant quantum computation'',
quant-ph/9712048.

\item{[\Freedman]}
M. H. Freedman,
``Percolation on the projective plane'',
\MRL\ {\bf 4} (1997) 889--894.

\item{[\Steanesimple]}
\steane,
``Simple quantum error correcting codes'',
\PRA\ {\bf 54} (1996) 4741--4751.

\item{[\QECCequiv]}
P. Shor and R. Laflamme,
``Quantum analog of the MacWilliams identities for classical coding
  theory'',
\PRL\ {\bf 78} (1997) 1600--1602;\hfb
\rains,
``Quantum weight enumerators'',
\IEEETIT\ {\bf 44} (1998) 1388--1394.

\item{[\Rainspoly]}
\rains,
``Polynomial invariants of quantum codes'',
quant-ph/9704042.

\item{[\Shornine]}
P. W. Shor,
``Scheme for reducing decoherence in quantum computer memory'',
\PRA\ {\bf 52} (1995) R2493--R2496.

\item{[\Gottesmanthesis]}
\gottesman,
{\sl Stabilizer Codes and Quantum Error Correction},
Caltech Ph.D. thesis, physics (1997),
quant-ph/9705052.

\item{[\five]}
C. H. Bennett, D. P. DiVincenzo, J. A. Smolin and W. K. Wootters,
``Mixed-state entanglement and quantum error correction'',
\PRA\ {\bf 54} (1996) 3824--3851;\hfb
R. Laflamme, C. Miquel, J. P. Paz and W. H. Zurek,
``Perfect quantum error correction code''.
\PRL\ {\bf 77} (1996) 198--201.

\item{[\CSSseven]}
A. R. Calderbank and P. W. Shor,
``Good quantum error-correcting codes exist'',
\PRA\ {\bf 54} (1996) 1098--1105;\hfb
\steane,
``Multiple particle interference and quantum error correction'',
\PRSLA\ {\bf 452} (1996) 2551--2576.

\item{[\Kitaevprivate]}
A. Kitaev, 
discussions on 21 September 1998.

\item{[\Berger]}
M. Berger,
``{\it Du c\^ot\'e de chez Pu\/}'',
\ASENSfour\ {\bf 5} (1972) 1--44.

\item{[\inprep]}
M. H. Freedman and D. A. Meyer,
``Systolically free quantum codes'',
in preparation.

\bye